%% file: ms.tex
\documentclass[sigconf]{acmart}

\usepackage{xcolor}
\usepackage{amssymb}
\usepackage{algorithm}
\usepackage{algorithmic}
\usepackage{multirow}
\usepackage[utf8]{inputenc}

\input{mathnotation.tex}

\def\full{0}       		
\def\shownotes{0}   	

\ifnum\shownotes=1
	\newcommand{\authnote}[2]{{ $\ll$\textsf{\footnotesize #1 notes: #2}$\gg$}}
	\newcommand{\authadd}[2]{{\blue{ \textsf{\footnotesize #2}}}}
	\newcommand{\authdel}[2]{{ \blue{\textsf{\footnotesize \sout{#2}}}}}
\else
	\newcommand{\authnote}[2]{}
	\newcommand{\authadd}[2]{#2}
	\newcommand{\authdel}[2]{}
\fi

\newcommand{\AMnote}[1]{{\red{\authnote{Arash}{#1}}}}
\newcommand{\AMadd}[1]{{\authadd{Arash}{#1}}}
\newcommand{\AMdel}[1]{{\authdel{Arash}{#1}}}

\newcommand\red[1]{\color{red}#1}

\newcommand\blue[1]{\color{blue}#1}

\newcommand{\system}{an information market}

\providecommand{\ie}{\emph{i.e.,} }
\providecommand{\eg}{\emph{e.g.,} }

\providecommand{\etal}{\emph{et al.}}   
\providecommand{\etc}{\emph{etc.}}      

\providecommand{\myparab}[1]{\smallskip\noindent\textbf{#1} }

\newcommand{\cpm}{\textrm{\texttt{CPM}}}
\providecommand{\ron}{\textrm{\texttt{RON}}_a}
\providecommand{\tqm}{\textrm{\texttt{TQM}}_p}
\newcommand{\intent}{\textrm{\texttt{Int}}}
\newcommand{\impl}{\textrm{\texttt{Impl}}}
\newcommand{\expl}{\textrm{\texttt{Expl}}}

\setcopyright{none}
\renewcommand\footnotetextcopyrightpermission[1]{}
\begin{document}

 \author{Arash Molavi Kakhki}
\orcid{1234-5678-9012}
\affiliation{
  \institution{ThousandEyes}
  \city{San Francisco, CA} 
  \state{USA} 
}
\email{arash.molavi@gmail.com}

\author{Vijay Erramilli}
\orcid{1234-5678-9012}
\affiliation{
  \institution{Salesforce}
  \city{San Francisco, CA} 
  \state{USA} 
}
\email{evijay@gmail.com}

\author{Phillipa Gill}
\orcid{1234-5678-9012}
\affiliation{
  \institution{University of Massachusetts}
  \city{Amherst, MA} 
  \state{USA} 
}
\email{phillipa@cs.umass.edu}

\author{Augustin Chaintreau}
\orcid{1234-5678-9012}
\affiliation{
  \institution{Columbia University}
  \city{New York, NY} 
  \state{USA} 
}
\email{augustin@cs.columbia.edu}

\author{Balachander Krishnamurthy}
\orcid{1234-5678-9012}
\affiliation{
  \institution{AT\&T Labs - Research}
  \city{New York, NY} 
  \state{USA} 
}
\email{bala@research.att.com}

 \title{Information Market for Web Browsing: Design, Usability and Incremental Adoption}

\begin{abstract}
	{\input{abstract}}
\end{abstract}

\keywords{Browsing privacy, Information market}

\maketitle
\input{introduction-augustin}

\input{background}

\input{system}

\input{user-study}

\input{analysis}
\input{relwork}
\input{conclusions}
\input{acknowledgment}

\bibliographystyle{abbrv}
\bibliography{PersonalData,CoopGameTheory,DoNOTEditAugustin}

\appendix

\input{appendixData}

\input{appendixShapley}

\end{document}

%% file: mathnotation.tex
\newcommand{\midwor}[1]{\;\textnormal{ #1 }\;} 
\newcommand{\ds}{\displaystyle} 
\newcommand{\vf}{\,,\,} 
\newcommand{\lset}{\left\{\left.\;}

\newcommand{\rset}{\;\right.\right\}}

\newcommand{\Real}{\mathbb{R}} 

\newcommand{\ind}[1]{\mathbb{I}_{\{#1\}}} 

\newcommand{\setA}{\mathcal{A}}

\newcommand{\setN}{\mathcal{N}}
\newcommand{\setS}{\mathcal{S}}

\newenvironment{disarray}%
 {\everymath{\displaystyle\everymath{}}\array}%
 {\endarray}

%% file: abstract.tex
Browsing privacy solutions face an uphill battle to deployment. Many operate counter to the economic objectives of popular online services (\emph{e.g.}, by completely blocking ads) and do not provide enough incentive for users who may be subject to  performance degradation for deploying them. In this study, we take a step towards realizing a system for online privacy that is mutually beneficial to users and online advertisers: an \emph{information market}. This system not only maintains economic viability for online services, but also provides users with financial compensation to encourage them to participate.  We prototype and evaluate an information market  that provides privacy and revenue to users while preserving and sometimes improving their Web performance. We evaluate feasibility of the market via a one month field study with 63 users and find that users are indeed willing to sell their browsing information. 
We also use Web traces of millions of users to drive a simulation study to evaluate the system at scale. We find that the system can indeed be profitable to both users and online advertisers.

%% file: introduction-augustin.tex
\section{Introduction} 
\label{sec:intro}

Most online services currently provide users with the same standard terms of usage: users may use them free of charge but those uses are monetized through data collection used for online advertising. This economic model was long criticized by privacy advocates, due to the growing amount of information collected about each user~\cite{Gill:2013,Krishnamurthy:2010ta}, and the lack of effective control it provides her over how it is used. Indeed, this model recently came under a new stress: Softwares that block ads and/or limit some third party tracking~\cite{adblock,noscript,dnt} are no more confined to a minority of savvy users: Adoption by Internet users is already reported up to 35\% in two European countries in 2015, and more importantly, quickly growing everywhere, including the largest advertising markewith the same standard terms of usage: ts~\cite{Anonymous:2015wp}. Some ad-blocking even comes shipped for the default browser on mobile platforms such as Apple iOS 9~\cite{ios-adblock}. Those users who \emph{in effect} opt-out of tracking or advertising (at least partially) are estimated to negatively affect ad revenues in the tens of billions~\cite{Anonymous:2015wp}. Those revenue losses motivated multiple research prototypes and development for alternative ways to collect, manage and exploit personal data: personal data store, lockers, intent-casting and privacy preserving ad personalization, see \cite{Anonymous:bkd81APP,Mun:2010uz,deMontjoye:2014bs} and more than 50 related development efforts mentioned at \texttt{cyber.law.harvard.edu/projectvrm/}.

Online privacy solutions such as those mentioned above typically struggle with restricted deployment due to the incentives (or lack thereof) that they produce. On the one hand a unilateral opt-out solution - deployed by a user herself, such as blocking~\cite{adblock,noscript} - may be difficult to configure as it often impacts service quality~\cite{cranor-adblock}. 
The incentive actually exists for ad-networks, aggregators, and publishers to cause disruption in service for such opt-out users, as its adoption by a user significantly reduces the value she generates ~\cite{Aziz:uz,Gill:2013ts}. 
In contrast, a cooperative solution - deployed jointly by multiple parties (\eg aggregators, publishers, users)~\cite{Guha:2011kn,Toubiana:2010tm} - may in principle offer incentives to all. However, most of these systems offer users enhanced privacy but no more, and little is formally known about the incentives offered to other parties. Much evidence suggests that enhanced privacy alone can be difficult for users to perceive and treat as a rational choice, even among users self-reporting a concern for tracking~\cite{Acquisti:2009tb} (a trend we confirm in our experiment). 

Here we evaluate through experiments and data-driven analysis the promise of a different online privacy solution: \emph{an information market}. In contrast to all the above systems, an information market implements an incentive for all parties to participate. Users, for instance, can enjoy enhanced privacy by selecting which of their data can be used and be compensated for those. Depending on how the information market is designed, other parties may find it profitable to use for their interaction with online users. The idea of information markets is not new~\cite{Laudon:1996jj,Ghosh:2015el,Cvrcek:2006vv,Riederer:2011ta}, and is even offered by some products today~\cite{datacoup,blackbox,socialdatacollective}\footnote{For simplicity we focus on financial incentives here, but compensation can be in the forms of upgrade or discounts.} but very little is known of their economic viability. Even less is known about how they should be designed to engage users, let alone how they could gradually encourage an incremental adoption. This paper provides a much needed evaluation of each of those aspects, concluding that information markets show some promise. While information markets could be applied to multiple forms of personal data, in this work we focus on their application to third party tracking during web browsing, for two reasons: First, third party tracking is a critical component of today's online advertising; second, the sharing of browsing data with third parties is often cited first among causes of concern by online users~\cite{Anonymous:2015vw}.

This paper presents the following contributions
\begin{itemize}
\item We design and build a simplified information market for third party web tracking. Through a simple architecture, this system enables selective privacy protection and economic transactions over data. Our design seamlessly integrates with today's web tracking and ad-network functions to be backward compatible. In fact, it even allows users to benefit from a performance boost as we show it can easily be combined with web acceleration features. 
(\S\ref{sec:system})
\item We conduct a 30-day experiment with 63 participants in two metropolitan areas to observe the effect of economic incentive on online users' behaviors. We confirm that self-reported privacy attitudes may not always align with actual data disclosures by users. More importantly, we validate that the two features of an information market, \ie privacy when you want it, compensation when you do not, are effectively used and managed by users. Every user has some data they choose to protect, although disclosing it would increase their earning up to 52\%. As their browsing expands and new data gets created, users also often choose economic returns over data protection. In fact, we observe a growing engagement during our user study, and an overwhelming majority claims to be likely to use such a system if it was deployed. 
(\S\ref{sec:experiment})
\item We study the potential for an incremental deployment of an information market at a large scale. To do so, we extend a model predicting today's online advertising revenue per user. Through cooperative game theory we analyze how market forces may affect revenue redistribution to different parties should information market be offered as an option. We then assume that a party adopts an information market only when positive benefits are gained, which may in turn affect others' decision to adopt. We use this model to analyze spread of adoption in several HTTP traces containing up to 3 million users, along with the publishers and third parties they interact with. This model predicts significant revenue growth (up to 9-12\%) as information market expands the availability of high quality data about users. A significant fraction of the users (from 35\% to 92\% depending on the traces) adopt information market and receive monetary gain.
(\S\ref{sec:deployment})
\end{itemize}

To our knowledge these results present the first data driven evaluation of information markets and their potential to scale to the web. We recognize that deployment of an information market remains a hard problem: Online user privacy is increasingly more complicated, as new forms of tracking or re-identification~\cite{Acar:2014bg} come into play. The complexity of the online advertising ecosystem creates multiple frictions to deployment\cite{Yuan:2012th}. Moreover, there is economic incentives for users to game the system (we monitored our user study closely and confirm we did not observe any instances of gaming to pollute our data and findings, and discuss in \S\ref{subsec:result} how gaming can be detected). Given those limitations, it is important to interpret our results in their context: First, they assume that the information market operates in a way that allows users to technically opt-out. 
Also our architecture allows the privacy preserving module to evolve to handle new forms of tracking and re-identification. 
Second, our analysis of online revenue and its redistribution necessarily makes some assumptions (following bargaining resolution as classically drawn by Nash and later Shapley). Our results prove at least that a few conventional wisdoms about online privacy may not always hold. Solutions offering privacy choices and compensating users for data may not necessarily reduce overall advertising revenue; they may even  benefit publishers. This contributes to the ongoing debate over who benefits from targeted advertising~\cite{Marotta:vi}. Finally, in terms of design, our results help in proving the promise of enabling selective privacy: our experiments show that users are able to determine the data they want to protect, and those they are eager to sell. For the latter, our trace analysis revealed for the first time that opportunities abound for an information market to make data more widely available, fueling more revenue. These trends may be reproduced under different conditions and assumptions than ours. Our findings contribute to argue that the arms race we experience between blocking and more invasive tracking techniques may not necessarily in the long term serve the interests of online publishers and the advertising industry overall.


%% file: background.tex
\section{Background and requirements}
\label{ref:sysreq}

Targeted advertising has increased in usage over the last few years and generally comes in multiple varieties: contextual, retargeting and behavioral~\cite{Liu:2013}. The last one has been offered by Google since 2009~\cite{googlead}. While contextual advertising serve ads based on the content of the page embedding the ads, both retargeting and behavioral utilize the browsing history of a user to place relevant ads. 

Past browsing behavior is obtained by aggregators via embedding themselves in Web pages as \emph{3rd-parties} in combination with setting cookies in the browser (NB: Other techniques such as fingerprinting exist; we discuss them in Sec.~\ref{subsec:ppp}). Consider the following example: (i) user Alice visits publisher \texttt{pubA.com} which contains references to a 3rd-party aggregator: \texttt{agg.com}.
(ii) If Alice is visiting \texttt{pubA.com} for the first time, the HTTP response is the content of the page being requested along with 
a \texttt{Set-Cookie} HTTP header, with cookies pertaining to \texttt{pubA.com}. Likewise, if Alice has never visited \texttt{agg.com} (or any page with them as a 3rd-party), the responses from this domain will also include \texttt{Set-Cookie} header.
(iii) Alice next visits publisher \texttt{pubB.com} that also contains references to aggregator \texttt{agg.com}. As a cookie for \texttt{agg.com} was set when she visited \texttt{pubA.com}, Alice's browser sends this cookie to  \texttt{agg.com}  along with its request. Thus, \texttt{agg.com} now knows that Alice visited both  \texttt{pubA.com}, and \texttt{pubB.com} and can use this to customize which ads to show. 

This type of 3rd-party tracking is considered objectionable by many, because it results in Alice's browsing history being revealed to 3rd-parties that she is unaware of. While publishers themselves may also track Alice over time -- 1st-party tracking -- Alice is generally aware of 1st-parties that she is dealing with. In contrast, a publisher may embed any number of 3rd-parties without notifying the user. 
\emph{The first goal of our system is to give transparency and control to Alice in the context of 3rd-party tracking.} 
A key difference with previous solutions to that problem is that Alice - aware that part of her browsing history may boost the revenue of the ads shown on the websites she visits - may sometimes be eager to disclose some of that information \emph{selectively} for an appropriate reward. \emph{Permitting such a data transaction is the second goal of our system.}

Overall, we hope to design an information system that can be built and used experimentally. It should hence satisfy the following requirements:


\begin{enumerate}
\item \textbf{Selective Privacy protection.} 
The system should protect users' browsing history from being revealed to 3rd-party aggregators. Note that this protection is critical even for the data the user intends to sell. Without privacy protection there is no incentive for aggregators to enter the market as they can obtain users' data via conventional means. The system should generally enable users to disclose only a fine-grained subset of their browsing.
\item \textbf{Backwards and incentive compatibility.} First, this means that the system should work with today's tracking and online advertising systems, with minimal modifications required for the delivery of ads. This helps in reducing friction in adoption for data aggregators and ad-networks. Second, it should encourage adoption by providing different parties with the right economic incentives. Note that this last requirement is more complex as decisions of several parties together interact to change revenue. This is why we will carefully study how different designs affect online revenue sharing.
\item \textbf{Access to data.} Unlike existing proposals~\cite{Chen:2013, Guha:2011kn,Toubiana:2010tm} {\system} should not require aggregators to communicate targeting algorithms--which may be considered trade secrets--to the system. Instead, aggregators should be able to purchase raw data about users' browsing habits (\ie when and what sites the user has visited). 
\item \textbf{Avoid having users price data.} Since aggregators have knowledge of the relative value of user data  (\eg somebody who visits Rolex.com may have higher value) they are in the best position to determine prices in the market. Further, having users assign value to their own 
data~\cite{Carrascal:2013} or calculate the loss of utility with their information release~\cite{Ghosh:2011jy} is non-trivial. We design an information market where the price is set by the demand for data from aggregators; we believe they are in a better position to price the data appropriately.

\end{enumerate}

%% file: system.tex
\section{System Description}
\label{sec:system}

\begin{figure}[t]
	\begin{center}
		\includegraphics[width=0.47\textwidth ]{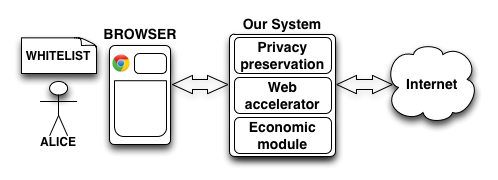}
		\caption{Overview of our system. Alice opts in and creates a whitelist, comprising of sites she's willing to disclose her visit to. Alice's Web requests are routed through our system. See Sections~\ref{subsec:ppp}-\ref{subsec:econ} for detailed descriptions for each module shown.}
		\label{fig:sys_overview}
	\end{center}
\end{figure}

\begin{figure}[t]
	\begin{center}
		\includegraphics[width=0.47\textwidth ]{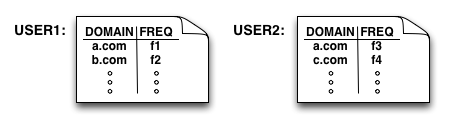}
		\caption{Sample of whitelists along with visit frequencies shown to aggregators to valuate users and bid on them. Note that aggregators will not know who users 1 and 2 are until they bid on them and win the auction on these users. They are then granted \emph{access} to these users and can track them on their whitelisted sites.}
		\label{fig:whitelists}
	\end{center}
\end{figure}

\begin{figure*}[t]
	\begin{center}
		\includegraphics[width=1\textwidth ]{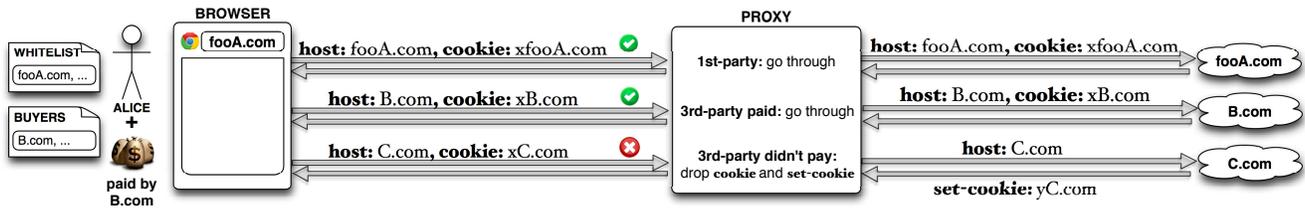}
		\caption{Alice visits a whitelisted site \texttt{fooA.com} with two trackers \texttt{B.com} and \texttt{C.com}. Since \texttt{B.com} has paid to access Alice, it receives cookies. \texttt{C.com}'s cookies are dropped in both directions. Note \texttt{B.com}'s cookies would be dropped if \texttt{fooA.com} is not whitelisted.}
		\label{fig:system}
	\end{center}
\end{figure*}

Our market solution, designed for the aforementioned requirements, is deployed in the network by a trusted third party \eg a government body or an ISP, as illustrated in Fig. \ref{fig:sys_overview}. First, we implement a privacy preservation module to keep user's browsing history private from 3rd-party aggregators. Users who opt-in to the market have all their Web requests routed through our system, which provides privacy protection by blocking known forms of Web tracking (Sec.~\ref{subsec:ppp}). Note that we do not block ads (or any other requests) and allow them to be shown to the user. The system only prevents leakage of personally identifiable information (PII) to advertisers.

Each user then creates a \emph{whitelist}, which is a list of Web sites, if any, they are willing to be tracked on by 3rd-party aggregators in return for monetary compensation. Users' whitelists along with their frequency of visits to those sites are then anonymized and presented to participating aggregators. Fig. \ref{fig:whitelists} shows a sample of what aggregators see. Aggregators can then valuate each anonymized user based on their whiteliste and frequency of visits to these sites, and bid on them if interested (Apx. \ref{apx:eval}). The system then runs an auction (Sec. \ref{subsec:econ}) for each user to determine the winning aggregators, if any, and the winning bid. Each winning aggregator then pays that user the winning amount and in return gets \emph{access} to the user for the duration of the next auction period. We define access as the ability to track the user across their whitelisted sites and use this information as input to any proprietary targeting algorithms to serve targeted ads to the user on those sites.. Note that adding a Web site to the whitelist by itself does not allow aggregators to track the user on that site. Aggregators still have to bid, win, and compensate the user before they can track them.

Next, we discuss each module in more detail.
\begin{enumerate}
	\item \textit{Privacy preserving module:} handles all web requests of all users and provides privacy.
	\item \textit{Web acceleration module:} compensates for any performance loss due to the added privacy protection.
	\item \textit{Economic module} handles auctions and economic transactions between the users and aggregators.
\end{enumerate}
\vspace{0.01in}

\subsection{Privacy preserving module}
\label{subsec:ppp}
The privacy preserving module handles all HTTP requests and responses for users with the objective to provide privacy protection 
against 3rd-party tracking, \emph{without} blocking any requests, including advertisements. Below we discuss different tracking methods and how the privacy preserving module deals with each of them:

\myparab{Cookies:}HTTP cookies are the most prevalent method for online tracking. Using cookies, a tracker can assign a unique identifier 
to each user, which is consistent across different websites, and gives them the power to track users browsing behavior. Today's browsers ship with options to block 3rd-party cookies. However, blocking 3rd-party cookies will not necessarily prevent 3rd-party tracking since most browsers only block \texttt{Set-Cookie} in HTTP responses, meaning if a 3rd-party already has its respective cookies set, \eg via popups as a 1st-party, then these previously set cookies will indeed be sent out by the browser to the 3rd-party tracker, defeating the purpose~\cite{Roesner:2012}. 
To prevent unwanted 3rd-party tracking via cookies, we first classify each HTTP request as either \textit{1st-party-request} or \textit{3rd-party-request}. A HTTP request is a 1st-party-request if the \texttt{Host} and \texttt{Referer} headers belong to the same root domain, and a 3rd-party-request otherwise~\footnote{With some exception cases, such as CDNs, which are treated as 1st-party. Our technique also wrongly classifies the following as a 1st party request: when a \texttt{publisher.com} uses CNAME to set the DNS to point part of their traffic, such as \texttt{metrics.publisher.com} to a 3rd party.}. Next, if 
a request is marked as a 3rd-party-request, we will only let its cookies go through along with the request if both of the following 
conditions are satisfied: 
\begin{description}
\item[ ] (a) the user making this request has whitelisted the root domain of the \texttt{Referer},\ie the top domain the user is visiting in their browser,
\item[ ] (b) the root domain of the \texttt{Host}, \ie the 3rd-party destination of the request, has paid to access this user (Sec.~\ref{subsec:econ}).
\end{description}

Cookies are stripped off from the 3rd-party-request and its response otherwise (Fig.~\ref{fig:system}). \texttt{Set-Cookie} per-se does not leak any PII, but it is dropped to preserve user's cookie jar consistency. Note that since we only strip information off of requests and don't block them, ads would still be rendered properly in the browser.


\myparab{Referers:}Referer headers may contain unnecessary PII~\cite{Krishnamurthy:2011uv}, and we treat them the same way as cookies, \ie drop if 
3rd-party-request and one or both of the conditions mentioned above are not satisfied.

\myparab{Web bugs:} 1x1 pixel bugs on a page are invisible to the user and serve no content. Their sole purpose is user tracking. The prior steps render these useless.

\myparab{\texttt{Etag} and \texttt{If-None-Match}:}These optional headers are generally used for web cache validation but can also be used for tracking~\cite{Krishnamurthy:2011uv}. 
When a web server is responding to a request, it can send an \texttt{Etag} header that is an identifier for the version of the resource being requested. 
When the user makes the request to the same resource in the future, she can send a conditional request using the \texttt{If-None-Match} header. 
Instead of a resource identifier, \texttt{Etag} can contain a unique user identifier for all new users, \ie users with no \texttt{If-None-Match} header in 
their request. When the same user returns to the web server, they identify themselves by sending the unique identifier 
as \texttt{If-None-Match} header, hence enabling tracking. We drop \texttt{Etag} and \texttt{If-None-Match} headers\footnote{The web acceleration module (Sec~\ref{subsec:web}) mitigates any possible bandwidth overhead to the web server.}.

\myparab{Flash cookies and LocalStorage:}Flash cookies and LocalStorage can be used to respawn deleted cookies~\cite{flash}. 
However the way we deal with 3rd-party cookies, renders respawning useless.

\myparab{JavaScript:}
JS can be used instead of \texttt{Set-Cookie} header. We already deal with this in the same way we deal with cookies.
JS can be used by 3rd-parties to collect information about users, \AMadd{using techniques such as fingerprinting~\cite{Anonymous:2013hd}}. Such code can be fingerprinted and blocked. This may be included in future versions of the system, but is beyond the scope of this paper.

\subsection{Web acceleration module}
\label{subsec:web}
To mitigate overhead of the privacy protection, we include a Web acceleration module. \AMdel{We implement a standard Web acceleration module, similar to Varnish~\cite{varnish} with some modifications. }To boost performance we use standard acceleration methods such as performing on-the-fly prefetching of static objects (images, JS, css), prioritizing traffic, image compression, and maintaining persistent TCP connections with both the user and the origin server and using the same connection for multiple requests.

\begin{figure}[t]
	\begin{center}
		\includegraphics[width=0.47\textwidth ]{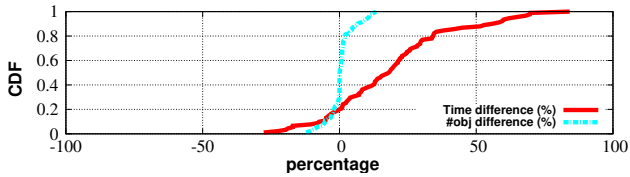}
		\caption{Performance and functionality of web acceleration and privacy preservation module. Positive percentages mean we are doing better.}
		\label{fig:awazza_performance}
	\end{center}
\end{figure}

\myparab{Benchmarks:} In order to test our web acceleration and privacy preserving modules for performance and functionality, we wrote a script using PhantomJS~\cite{phantomjs} and ySlow~\cite{yslow} to load each Alexa top 100 sites once via our system and once directly. We repeat this test 10 times and look at the average difference in load time and number of objects as a measure of functionality\footnote{The expectation is that privacy protection mechanisms do not block any legitimate content that can lead to lower quality of experience for the end-users.}. Fig.~\ref{fig:awazza_performance} shows that roughly 80\% of sites load faster with our accelerator. With regards to the number of objects, while most sites have the exact same number, some have fewer or more objects when loaded via our system. This is an expected result since most pages are dynamic and the content does not necessarily remain the same over the course of the experiment. We manually inspected pages with different numbers of objects to confirm that the difference is mostly due to content change.


\subsection{Economic module}
\label{subsec:econ}



At the beginning of each auction period, the economic module does the following:
\begin{enumerate}
	\item Presents aggregators with \emph{anonymized} whitelists and frequency of visits.
	\item Collects aggregators bids on users. Aggregators bid independent from each other.
	\item Runs an auction per user and determines the winning value and winning aggregators (if any) for each of the users. Note that a user's browsing behavior can be sold to multiple aggregators.
	\item Handles transactions between users and aggregators and makes sure every aggregator pays all the users they won access to. It then notifies the privacy preserving module to grant appropriate access for the duration of the next auction period.
\end{enumerate}

We let aggregators compute the value of access to a user for the following reasons. First, aggregators have experience deriving value from PII. Second, they are able to assess revenues on a short-term basis through the sale of goods or ad-space, compared to the long-term risk a user must calculate in dealing with privacy.  Finally, aggregators deal with many customers, and can take more risk in over- or underestimating the value of access, as opposed to users who are more risk-averse.
However, there can be strong incentives for aggregators to lie about their valuation by under-bidding, which results in lowering the user's revenue. To prevent such behavior, the auction should have a truth telling mechanism where aggregators gain no additional benefit by under-bidding. 
For this purpose, we rely on an auction mechanism called the exponential mechanism~\cite{exp-mechanism} that has truth telling properties, and has been shown to be close to optimal in terms of revenue for the seller (user in our case)\footnote{We choose this latter objective, while noting that other objective functions (\emph{e.g.},~maximizing revenue for all players in the value chain) can be chosen.}. Apx. \ref{app:auction} explains the auction mechanism in detail.

%% file: user-study.tex
\begin{table*}[t]
\begin{center}
	\small
	\centering
		\begin{tabular}{ c | p{10cm}  | c|c|c|c|c}
			& \multicolumn{1}{c|}{Question} & \multicolumn{5}{c}{Answers} \\
			& & 5 & 4 & 3 & 2 & 1 \\
			\hline
			\multirow{8}{*}{Pre} & \textbf{1-Are you concerned about protection of your private data online?*} & 21\% & 19\% & 47\% & 11\% & 2\%\\
			& \textbf{2-Do you read the privacy policies of the websites you visit?*} & 	0\% & 3\% & 27\% & 37\% & 33\%\\
			& \textbf{3-Do you use private browsing/Incognito mode while browsing?*} & 0\% & 14\% & 36\% & 40\% & 10\%\\
			& \textbf{4-Do you let mobile applications access your location on smartphone?*} & 5\% & 25\% & 36\% & 27\% & 7\%\\
			& \textbf{5-Do you maintain several passwords for multiple websites?*} & 13\% & 41\% & 38\% & 8\% & 0\%\\
			& \textbf{6-Do you think ads you see online are becoming more relevant to you?*} & 14\% & 27\% & 35\% & 21\% & 3\%\\
			& \textbf{7-How much are you willing pay for privacy protection?**}  & 2\% & 3\% & 14\% & 49\% & 32\%\\
			& \textbf{8-How much do you value today's free online services?**} & 16\% & 14\% & 29\% & 35\% & 6\%\\
			\hline
			\multirow{13}{*}{Post} & \textbf{1-How comfortable were you to receive micro-payments?*}  & 39\% & 34\% & 17\% & 8\% & 2\%\\
			& \textbf{2-How satisfied are you with the amount you got for you information?*} & 17\% & 50\% & 18\% & 13\% & 2\%\\
			& \textbf{3-How satisfied are you with the performance of the system?*} & 20\% & 45\% & 27\% & 5\% & 3\%\\
			& \textbf{4-Did system give you better transparency on which data about you is used?***} & 57\% & 32\% & 7\% & 3\% & 1\%\\
			& \textbf{5-Are you likely to use such a system if it was offered large scale?*} & 56\% & 20\% & 15\% & 7\% & 2\%\\
			& \textbf{6-Was the system fair in recognizing the value of your information?***} & 38\% & 42\% & 12\% & 8\% & 0\%\\
			& \textbf{7-Did such a system increase your concern about your personal data?*} & 13\% & 20\% & 27\% & 23\% & 17\%\\
			& \textbf{9-Did system increase awareness on how online services monetize your data?*} & 28\% & 38\% & 17\% & 12\% & 5\%\\
			& \textbf{10-Are you likely to pay for privacy now that study has ended?**} & 0\% & 0\% & 28\% & 33\% & 39\%\\
			& \textbf{11-Did using this system make you more likely to choose a paid service that comes with privacy guarantee over a free service?*} & 3\% & 7\% & 12\% & 35\% & 43\%\\
			& \textbf{12-How likely are you to pay for the online services that you use for free today after this study?*} & 3\% & 7\% & 15\% & 35\% & 40\%\\
		\end{tabular}
	\caption{Questions asked in our pre and post study questionnaires. Users answer on a 5-point Likert Scale: *(5: A lot, 1: Not at all), **(5:  >\$50, 4: \$20-\$50, 3: \$5-\$20, 2: \$0-\$5, 1: \$0), ***(5: completely agree, 1: completely disagree)}
	\vspace{-0.2in}
	\label{table:questions}
	\end{center}
\end{table*}

\section{User study}
\label{sec:experiment}

We implemented our system and performed a field study to understand user engagement, usability\AMdel{aspects of an information market}, and how users perceive economic rewards for their data. 

\subsection{Implementation}
\label{subsec:impl}
\AMnote{This subsection was moved from Section \ref{sec:userstud} to here}

The privacy preserving and web acceleration modules are managed by a proxy, hosted on Amazon EC2, that handles all requests and treats 3rd-party-requests according to the mechanisms described in ~\ref{subsec:ppp}. For each user, the proxy maintains a whitelist, as well as a \emph{buyers list} that includes aggregators who have paid to access that user.

\begin{figure}[h]
	\begin{center}
		\vspace{-0.1in}
		\includegraphics[width=0.2\textwidth ]{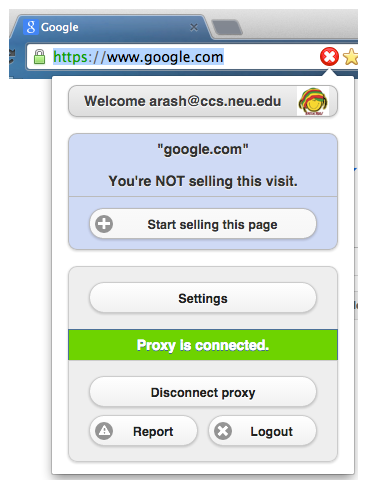}
		\vspace{-0.13in}
		\caption{Screenshot of our browser extension. User is visiting \texttt{google.com} which is not in their whitelist.}
		\vspace{-0.25in}
		\label{fig:extension}
	\end{center}
\end{figure}

We developed a browser extension to serve as an interface for users to interact with the market. Figure \ref{fig:extension} shows a screenshot of the extension. It places a color-coded icon in the address bar; green for whitelisted sites, red otherwise. Users can click to add/remove sites to/from the whitelist, or check their current whitelist, buyers list, and past earning.

\subsection{Experimental setting}
\label{subsec:eval}

\myparab{User study description}
A total of 63 participants located in New York City and Boston used our system for 30 days. Our population overrepresents male (75\%), and young (85\% under 25) users, and has wide variations in income (median around \$35k), technical proficiency and Internet use.

Every user was initially assigned a default whitelist containing 108 websites chosen from Alexa top 125 sites excluding social network and adult sites. Users could modify their whitelist at any time. Auctions, using the mechanism shown in Sec.~\ref{subsec:econ}, were conducted every three days, resulting in 10 data points per user. At each auction, the goods for sale are 
the ability to track users for the upcoming period.
Each winning aggregator pays the winning bid value to the user and gains access to her browsing activity \emph{on her whitelisted sites only} during the subsequent auction period to deliver targeted ads.
As an example, aggregator \texttt{agg.com} wins access to user $u$ with whitelist $WL_u$. This implies that, until the next auction, the privacy preserving module will NOT modify 3rd-party-requests to \texttt{agg.com} 
caused by $u$ while visiting sites in $WL_u$. 

We did not have access to real aggregators for our study, hence we \emph{emulated} 128 aggregators, including big players such as DoubleClick and Facebook. The challenge when emulating aggregators is computing their valuation for users' data.
For this we use a simple model based on keywords and their estimated cost-per-click commonly made available by ad-networks (details in Apx.~\ref{apx:eval}). Note that, based on the emulated valuation of their data and the auction's results, users were paid using \emph{real} money.


Each user was also asked to answer a questionnaire before and after the study, The pre-study questionnaire included demographical questions as well as self reported assessment of the user's view on online privacy. The post-study questionnaire contained questions about the user's experience with the system, and assessed if using our system changed her view on online privacy. Table~\ref{table:questions} shows the questions and answers we received.

\subsection{Users and information markets}
\label{subsec:result}
\AMnote{The structure of this section was changed to (1) state what we see (2) give supporting data that backs up what we see}


\AMadd{We made the following observations during the user study:}

\myparab{i) Users actively engage with the system and edit their whitelists.}
We first studied users' engagement with the tool and how they reacted to different facets of the system. 
We logged the number of HTTP requests coming from each user, which we refer to as \textit{activity}. 
The average activity stays in the [10k,14k] range per user during all 10 auction periods, which indicates users were using the system on a regular basis. Users actively added/removed sites to/from their whitelists, showing their engagement with the system, and we observed a sustained growth of the whitelists used, with an average user adding overall 50 new sites (a net 45\% increase from default whitelist) during the experiment.

As the post-study responses suggest Table~\ref{table:questions}, the majority of users were happy with the system's performance, receiving micro-payments, and the amount they were making for the data they released. 
\AMdel{We also observed a sustained growth of the whitelists used, with an average user adding overall 50 new sites (a net 45\% increase from default whitelist) during the experiment.}

\myparab{ii) Users' disclosure does not match up with their stated privacy attitudes.}
We looked into users' behavior with regards to data disclosure and how it correlates with their \emph{self-reported} privacy concerns \AMdel{(refer to Table \ref{table:questions} for question details and summary of answers)}. We took the sum of the 5-point Likert Scale answers that users gave to privacy related questions \AMdel{(numbered 1, 2, 3, 5, and 7)} in the pre-study survey. This gives a \textit{privacy score} for each user, ranging from 8 to 20, where higher scores indicates higher privacy concerns. Surprisingly, we found \emph{no} sign of negative correlation between users' privacy score and size of their whitelist; the associated correlation coefficient is low: $-0.1$. This, along with previous results~\cite{Acquisti:2009tb}, suggests that users stated privacy concerns are at odds with their behavior. One possible explanation for this is that users are swayed by the economic returns.

\begin{table}[]
	\small
	\centering
		\begin{tabular}{ c | c | c | c | c }
		Bucket & {earning} & {size of WL*} & {\#WL visits} & {WL/ALL} \\
		 & (USD) &  &  & {activity} \\
		\hline
		HD & 2.76	& 230.6 & 151.6 & 0.71\\
		\hline
		MD & 2.12 & 128 & 128.1 & 0.52\\
		\hline
		LD & 1.55	 & 115.1 & 67.4 & 0.48\\
	\end{tabular}
	\caption{averages of earning per auction, size of whitelist, number of whitelisted visits during auction period, and the fraction of activity which is whitelisted for different buckets. *WL = Whitelist}
	\vspace{-0.15in}
	\label{table:pri_money}
\end{table}

\myparab{iii) Users trade off their revenues to preserve privacy of their browsing behavior.}
To understand earning potentials of different users, we sort users based on their whitelists' size and split them into three equal size buckets: high, moderate, and low volume disclosers (resp. HD, MD and LD). Table~\ref{table:pri_money} shows earning, whitelist size, number of whitelisted visits, 
and fraction of whitelisted activity for each buckets, averaged across users and auction periods. HD users earn 31\%/79\% more than MD/LD users. A primary factor is that they have larger whitelists (80\%/100\% more than MD/LD) that results in 71\% of their activities coming from whitelisted sites, in contrast with 52\%/48\% for MD/HD users. 
If users were to disclose everything, \ie add \emph{all} the sites they browse to their whitelist, HD, MD, and LD users could potentially increase their earnings by 29\%, 48\%, and 52\% respectively. 
\AMdel{In other words, users may not follow what they initially self-report but a user do select a \emph{different} tradeoff between economic returns and privacy, and they leave money on the table when privacy is at stake.}
\AMadd{In other words, while users' behavior may not exactly follow their initial self-reported concerns, they\AMdel{ are willing to leave money for privacy} are willing to trade off their revenues to preserve privacy of their browsing behavior.}

\myparab{iv) Users pay attention to their earning and have incentive to game the system.}
\label{subsubsec:gaming}
\AMnote{Following crossed out paragraph might seem contradicting to previous observation we make and confuse the reader. It also doesn't really fit here.}
\AMdel{As Table~\ref{table:corr} shows, earning primarily correlates with whitelisted visits, that generates impressions with aggregators.
We saw high volume disclosers (HD) earning more, but the size of whitelist is not itself a great predictor: a user might have a large whitelist and small activity, or vice versa. 
As there are economic incentives in play, it is only natural to expect gaming by users -- users may decide to browse more as they expect economic returns. A key challenge is to differentiate between users who are abusing the system and users who are genuinely releasing more data, by choosing to be less private or not all.}
As there are economic incentives in play, it is only natural to expect gaming by users -- users may decide to expand their whitelists and browse more in order to increase their economic returns. A key challenge is to differentiate between users who are abusing the system and those who are genuinely releasing more data, by choosing to be less private and/or are more active than average users. 

To study this aspect, we sorted users based on their total earning in auctions 1 to 5, and divided them into three groups: top-earners (8 users), low-earners (8 users), and avg-earners (rest). On average top-earners made 6X more money than low-earners and we observed, as expected, top-earners have indeed bigger whitelists (36\% bigger than low-earners on average) and browse more (4.5X more than low-earners on average). 
To evaluate whether users would adjust their behavior and ``game'' the system to increase their earnings,
for the rest of the auctions, \ie auctions 6 -10, we gave top-earners a third of their earning and low-earners 3 times their earning. While this change did not have any significant effect on \AMdel{most users'} \AMadd{low-earners'} behavior -- pointing to normal behavior --  we did make a few interesting observations \AMadd{about top-earners}: 1) we received an email from one of the top earners complaining that the earning suddenly dropped even though their browsing habits had not changed 2) two top earners completely abandoned the system after the 7th auction, 3) three top-earners roughly doubled their whitelist \AMadd{trying to raise their earning back to previous range}, 4) one top earner doubled his/her browsing activity.
\AMdel{We did not observe behavioral changes for low-earners.} This suggests users pay attention to their earning, especially once it is significant, and that there is room for gaming the system\AMdel{, although none of these users are necessarily malicious}. Our experiment did not prevent gaming, it is however feasible in the future to include anomaly detection methods to detect abnormal activities. 

\myparab{v) Information market impacts privacy attitudes.}
Most users agree after the experiment that the system gives them better transparency \AMadd{(96\%), increases their awareness of data monetization (83\%),} and they claim to be likely to use it if it is deployed (91\%); see Table~\ref{table:questions}. The experiment did not noticeably increase their privacy concerns, nor did it affect substantially the \AMdel{\emph{median or average}} price they would pay for privacy. 
In other words, while users may still not pay for privacy, information markets did help them form a more consistent opinion about  their online privacy and value of their data.

\AMdel{However, we notice two interesting secondary effects: the distribution narrows as its extreme values noticeably drop, and users are less willing to pay for online services that are currently free. These findings suggest that users of information markets may not fear or even pay for their privacy but could form more consistent opinions about the value of their data.}

%% file: analysis.tex
\section{Incentives for Adoption}
\label{sec:deployment}

\begin{table*}[t]
	\begin{minipage}[c]{0.26\linewidth}
		\begin{center}
			\textbf{Mediated marketplace}
			\vspace{2ex}

		{\small
			\begin{tabular}[c]{c|c}
				$\setS$ & $\intent_a(u)$ 
				\\ \hline 
				$\emptyset$, 
				$\{ u \}$, 
				$\{ a \}$, 
				$\{ m \}$, & $\impl_a(u)$  
				\\
				$\lset u,a \rset$, 
				$\lset a,m \rset$ & $\impl_a(u)$  
				\\
				$\lset u,m \rset$ & $1$  
				\\
				$\lset u,a,m \rset$ & $\expl(u)$  
				\\
			\end{tabular}
			\begin{tabular}[c]{c|c}
				& Shapley value 
				\\ \hline 
				$u$, $m$ & $\frac{\expl(u)-1-\frac{3}{2}{(\impl_a(u)-1)}}{3}$ 
				\\
				$a$ & $\impl_a(u)+\frac{\expl(u)-1}{3}$ 
				\\
			\end{tabular}

			}
		\end{center}
	\end{minipage}
	\begin{minipage}[c]{0.26\linewidth}
		\begin{center}
			\textbf{Direct marketplace}
			\vspace{2ex}

		{\small

			\begin{tabular}[c]{c|c}
				$\setS$ & $\intent_a(u)$ 
				\\ \hline 
				$\emptyset$ & $\impl_a(u)$ 
				\\
				$\lset u \rset$ & $1$  
				\\
				$\lset a \rset$ & $\impl_a(u)$  
				\\
				$\lset u,a \rset$ & $\expl(u)$  
				\\
			\end{tabular}

			\begin{tabular}[c]{c|c}
				 & Shapley value 
				\\ \hline 
				$u$ & $\frac{\expl(u)-1-2(\impl_a(u)-1)}{2}$ 
				\\
				$a$ & $\impl_a(u)+\frac{\expl(u)-1}{2}$ 
				\\
			\end{tabular}

			}
		\end{center}
	\end{minipage}
	\begin{minipage}[c]{0.24\linewidth}
		\begin{center}
			\textbf{Publ. DNT mediated}
			\vspace{2ex}
			
		{\small

			\begin{tabular}[c]{c|c}
				$\setS$ & $\intent_a(u)$ 
				\\ \hline 
				$\emptyset$, 
				$\{ u \}$, 
				$\{ a \}$, 
				$\{ m \}$, & 1
				\\
				$\lset u,a \rset$, 
				$\lset a,m \rset$ & 1
				\\
				$\lset u,m \rset$ & $1$  
				\\
				$\lset u,a,m \rset$ & $\expl(u)$  
				\\
			\end{tabular}

			\begin{tabular}[c]{c|c}
				& Shapley value 
				\\ \hline 
				$u$, $m$ & $\frac{\expl(u)-1}{3}$ 
				\\ $a$ & $\frac{\expl(u)-1}{3}$ 
				\\ 
			\end{tabular}

			}
		\end{center}
	\end{minipage}
	\begin{minipage}[c]{0.22\linewidth}
		\begin{center}
			\textbf{Publ. DNT direct}
			\vspace{2ex}

		{\small

			\begin{tabular}[c]{c|c}
				$\setS$ & $\intent_a(u)$ 
				\\ \hline 
				$\emptyset$ & 1 
				\\
				$\lset u \rset$ & $1$  
				\\
				$\lset a \rset$ & $1$
				\\
				$\lset u,a \rset$ & $\expl(u)$  
				\\
			\end{tabular}

			\begin{tabular}[c]{c|c}
				 & Shapley value 
				\\ \hline 
				$u$ & $\frac{\expl(u)-1}{2}$ 
				\\
				$a$ & $\frac{\expl(u)-1}{2}$ 
				\\
			\end{tabular}

			}
		\end{center}
	\end{minipage}
	\caption{Various market types and coalition scenarios: coefficient multiplying advertising revenue for each coalition scenario (top), Shapley value for each party (bottom).}
	\vspace{-0.1in}
	\label{tab:revenue}
\end{table*}

An information market affects the revenue produced by advertisement overall and also how it is shared. We capture this effect first in a model focusing on the transaction associated with a single user. We then describe data used to feed this model for millions of users, and a simple adoption dynamics based on immediate financial incentive. Putting those together, we can then study the spread of adoption and the effect of information market at web scale. 

\subsection{Incentive for a single user}
\label{sec:econmarket}
Our model leverages recent modeling of online advertising~\cite{Gill:2013} to understand how advertising revenue changes with availability of information. Understanding how revenue is to be distributed among all parties is more difficult to predict. But one way is to apply the theory of cooperative games~\cite{Shapley:1953ty} which offers a principled approach to compute the expected outcome of a negotiation in the presence of alternative offers.

\subsubsection{Advertising revenue model and assumptions} 

Our model is inspired from prior work~\cite{Gill:2013} breaking down the value of ad-impression, or cost-per-mille (CPM), into a function of three factors:
\begin{equation}
\cpm(u,p,a)=\ron \times \tqm \times \intent_a(u) \vf
\label{eqn:cpm}
\end{equation}
where $\ron$ (run-on-network) is an ad's nominal cost in ad network $a$, $\tqm$ (traffic quality multiplier) captures the quality of the publisher $p$ (\eg reputable publishers vs. sites distributing copyright infringing content), and $\intent_a(u)$ (intent) depends on the value of information gleaned about the user $u$ by the ad network. 
Intent captures the fact that an impression can be sold for a larger price if the ad network knows that the user has performed some previous actions (\eg frequent or recent visits to a product related webpage). 

We include three cases for intent. 1) If the user chooses not to reveal any information, this coefficient is 
by convention equal to 1. 2) When tracking is not blocked, it takes on the value which we refer to as \emph{implicit} intent, $\impl_a(u)\geq 1$,
which depends on how much 
information the ad-network can collect about the user's browsing. 
3) If the user decides to release \emph{all} legitimate 
sites available to $a$, the impression can be sold with a higher \emph{explicit} intent $\expl(u)\geq\impl_a(u)$ 
independent of $a$. 


\myparab{Revenue sharing.} 
We treat the advertising transaction as a cooperative game capturing a simple dynamic: the more user information is known by the ad-network, if they cooperate via an information market, the larger the revenue. We assume that each player is incentivized by receiving a share of the revenue computed using the \emph{Shapley value}~\cite{Shapley:1953ty}. This mechanism has two advantages: it ensures some form of fairness, and\AMdel{ it typically lies in the \emph{core}, which} maximizes the likeness of cooperation, \ie users and aggregators participating in the market 
(see Apx.~\ref{apx:shapley} an overview of the Shapley value).

In practice, the aggregator $a$ typically collects payments from advertisers, gives a constant fraction $(1-\alpha)$ to the publisher, and then pays for data on the market at price set by this user's Shapley value. The market provider also receives its Shapley value from $a$. 

\subsubsection{Allocation of revenue to each player}
\label{subsec:alloc}
In this section we investigate the impacts on the Shapley value of different cooperative scenarios between players, \ie users, aggregators, and the market, under various market implementations. Table~\ref{tab:revenue} shows the intent, $\intent_a(u)$, which impacts the revenue that different cooperating subsets of players, $S$, can produce. 

We start with the situation in which a \textbf{mediated market} $m$ centralizes data collection and/or analytics as in~\cite{Saikat:2011tt,Riederer:2011}  (leftmost subtable in Table~\ref{tab:revenue}). Without any cooperation, as in today's status quo, a user is tracked implicity, $\intent_a(u)=\impl_a(u)$. However, if user and aggregator join the market, a higher revenue can be obtained corresponding to $\intent_a(u)=\expl_a(u)$. In one case, assuming $m$ and $u$ collaborate (\ie~user sells her data) but $a$ does not (\ie~$a$ decides not to buy it), revenue effectively decreases as implicit tracking is now blocked, $\intent_a(u)=1$.

The analysis of this game yields Shapley values for each player (shown in Table~\ref{tab:revenue} bottom). These values specify how much each player receives from the \emph{surplus}, \ie the potential revenue that is not produced today due to lack of cooperation, which is proportional to the added value of $\expl(u)$ over $\impl_a(u)$.
Based on Table 3 we make the following observations:

\myparab{i) Ad-networks are always better off buying the users' data.} Buying data makes them join the coalition, always leading to a revenue increase. Hence they always receive a positive Shapley value (Table~\ref{tab:revenue}).

\myparab{ii) Users are not always better off selling their data.} Precisely, a user is able to claim a positive share of the surplus only if this surplus is large enough to compensate for the blocking of implicit tracking. In a mediated market, it occurs as 
\(
\left(\expl(u)-1\right)-\frac{3}{2}(\impl_a(u)-1)>0
\)\(
\midwor{or} r_{a}(u)>\frac{3}{2} 
\),
where
\( 
r_{a}(u) = \frac{\expl(u) - 1}{\impl_a(u) - 1} 
\)
{is the \emph{consent tracking lift}.}


\myparab{iii) Mediated markets lower the bar for users to gain revenue from the market compared with direct markets.} Our analysis extends to different market implementations. While mediated markets simplify deployment, and offers a single point of sale to aggregators, other architectures run data collection and/or analytics on users' end~\cite{repriv,Toubiana:2010tm}. This creates a \textbf{direct market} with aggregators entering transactions with users directly (Table~\ref{tab:revenue} second from the left). All previous observations generalize except that users are better off only if 
$r_{a}(u)>2$. 

\myparab{iv) Publishers enforcing Do Not Track create ideal conditions to deploy an information market}.
We also analyze information markets deployed over sites that enforce Do Not Track (Table~\ref{tab:revenue}, third and fourth from left). In these scenarios publishers comply with regulations or aim to improve their image vis-a-vis privacy watchdogs and aggregators \emph{and} users are \emph{always} better off buying/selling data via the market. 

\myparab{v) As soon as the possibility of users adopting an information market is credible, a publisher always benefits in the long term by enforcing  DNT}. Note that publishers are driven by revenue, and since they receive a fixed share, they have an incentive to keep intent (and thus revenue) as high as possible. Enforcing DNT might initially undermine publishers' profit, however it does\AMdel{ redistribute claims to surplus in a way that} incentivize users to sell their data, making\AMdel{ intent and hence a} publishers' revenue higher in the long term.


Information markets increase ad-revenue and redistribute part of the surplus to all participating entities. The price of data set by the market always ensures that aggregators make a profit. However, it does not necessarily mean that a user is always better off selling their data. In addition, a user's decision to join a market is done only once, affecting all of its related ad-revenue. 

Here we apply our model to traces containing billions of Web requests made by millions of users to understand how information markets adopt at scale and the effect they have on total and individuals' revenue. 

\subsection{Data and adoption dynamics}

\myparab{Browsing data-sets} 
We use anonymized HTTP traces from a university network, a neighborhood of broadband users, and a country wide 
mobile ISP with approximately 8k, 5k, and 3M users, respectively. We process these traces into HTTP sessions and identify publishers and aggregators for each session, for each user. Intent values $\impl_a(u)$ and $\expl_a(u)$ are computed using browsing profiles observed with partial/global views (see~\cite{Gill:2013}). \AMdel{This is conservative as in practice, a market may facilitate the sale of more valuable data not readily mined from Web activity, increasing $\expl_a(u)$ and hence facilitating adoption.}

\begin{figure}[t]
\centering
\vspace{-1.1cm}
\includegraphics[width=0.4\textwidth ]{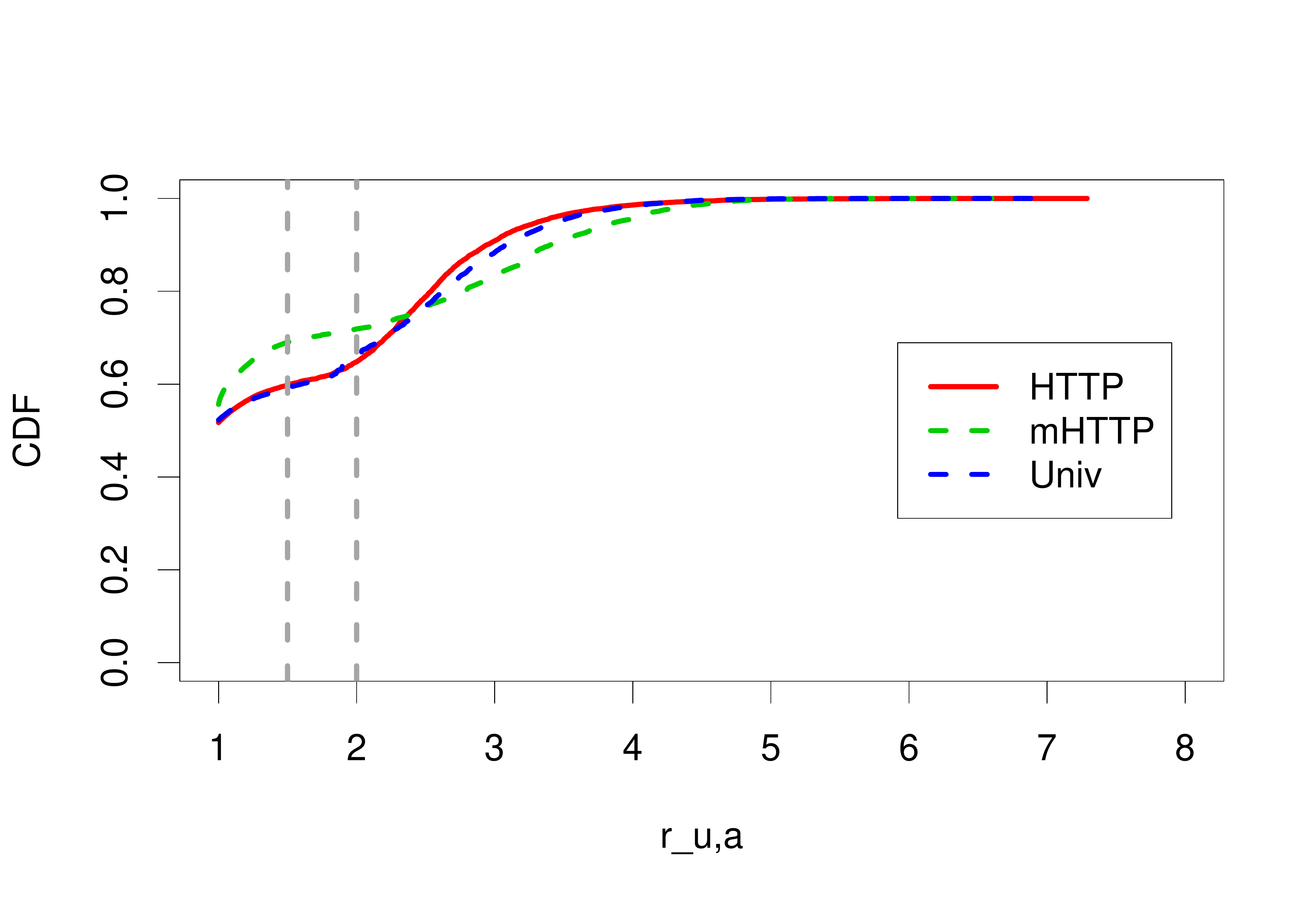}
\vspace{-0.2in}
\caption{Distribution of consent tracking lift $r_u(a)$, defined as the ratio of advertising value with explicit intent (obtained at user consent) divided by implicit intent (obtained from tracking).}
\vspace{-0.1in}
\label{fig:rua}
\end{figure}

\myparab{Distribution of consent tracking lift}. The increased value of data in the market, relative to what aggregators can infer ($r_{a}(u)$) is shown in Fig~\ref{fig:rua} for all user-aggregator pairs. 
Only 30-40\% of user-aggregator pairs have $r_{a}(u)$ above 1.5 (threshold for a user to sell data with a benefit in a mediated market as shown in Sec.~\ref{subsec:alloc}) and 28-34\% have $r_{a}(u)$ above 2 (same threshold for a direct market). At first glance, the surplus appears often too small for the information market to positively benefit users. This is especially true in the mHTTP dataset, where the aggregators appear to make accurate inferences, leading to high implicit intent which limits the value of $r_a(u)$.

\myparab{Myopic best response dynamics}
We assume each player aims to maximize its immediate profit as it decides whether to join an information market or not: \textbf{A user} will not join an information market unless there exists an ad-network for which it can claim positive profit (\ie  $r_{a}(u)>2$ for a direct market, or $r_{a}(u)>3/2$ for a mediated one). Note that by doing that, it also blocks tracking of other aggregators\AMdel{, but it does not pay the associated negative Shapley value since the ground to enforce it appears weak (users received currently no payment for this ad-revenue)}. \textbf{An ad-network} joins an information market 
as soon as joining the market will increase its revenue across all users and publishers in the ecosystem. Note that if many users have joined the information market (and blocked tracking) the aggregator may increase its revenue by joining the market, but still fall short of its initial revenue values before any market deployment. We analyze this in more detail in~\ref{subsec:effectonrev}. 


\subsection{Spread of adoption}

\begin{figure}[t]
\centering
\vspace{0.15in}
\includegraphics[width=0.44\textwidth ]{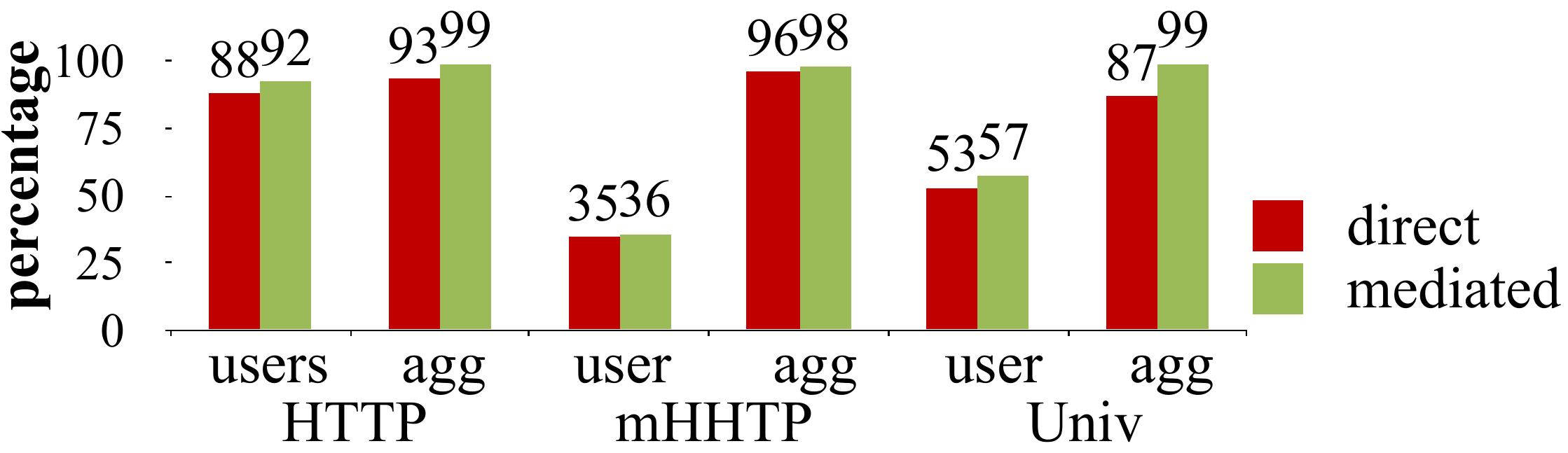}
\vspace{-0.1in}
\caption{Percentage of users and aggregators in the market at the end of the simulation.
\vspace{-0.19in}
}
\label{fig:fracusersagg}
\end{figure}

\ifnum\full=1
\begin{figure}[t]
\centering
\includegraphics[width=0.45\textwidth ]{graphs/aggsBuying.pdf}
\vspace{-0.7cm}
\caption{Fraction of aggregators in the market in each iteration.}
\label{fig:aggsbuying}
\end{figure}
\fi

Fig.~\ref{fig:fracusersagg} shows the percentage of users and aggregators that eventually join the market. In contrast with what our preliminary analysis of $r_u(a)$'s distribution predicted, we see a higher adoption rate, especially among aggregators. This shows that even a small number of pairs $(u,a)$ with high $r_u(a)$ is sufficient to generate a network effect of adoption. Indeed, more than 87\% of the aggregators purchase data from at least one user, and 14-22\% of those do so only because a user previously joined a market. As predicted, the mediated market has a higher percentage of adoptions with an increase of 4-8\% relative to the direct market.

\ifnum\full=1
Fig.~\ref{fig:aggsbuying} shows the fraction of aggregators that join the market in each iteration of the simulation.
\fi

The same effect, although not as pronounced, is true for users: more than {35\% and up to 92\% see a positive profit and hence sell their data.} Note that here we are only considering economic incentives for users to join, while some additional users may do so for other concerns such as privacy.

\subsection{Effects on ad-revenue}
\label{subsec:effectonrev}
\begin{figure}[t]
\centering
\vspace{0.25in}
\includegraphics[width=0.44\textwidth ]{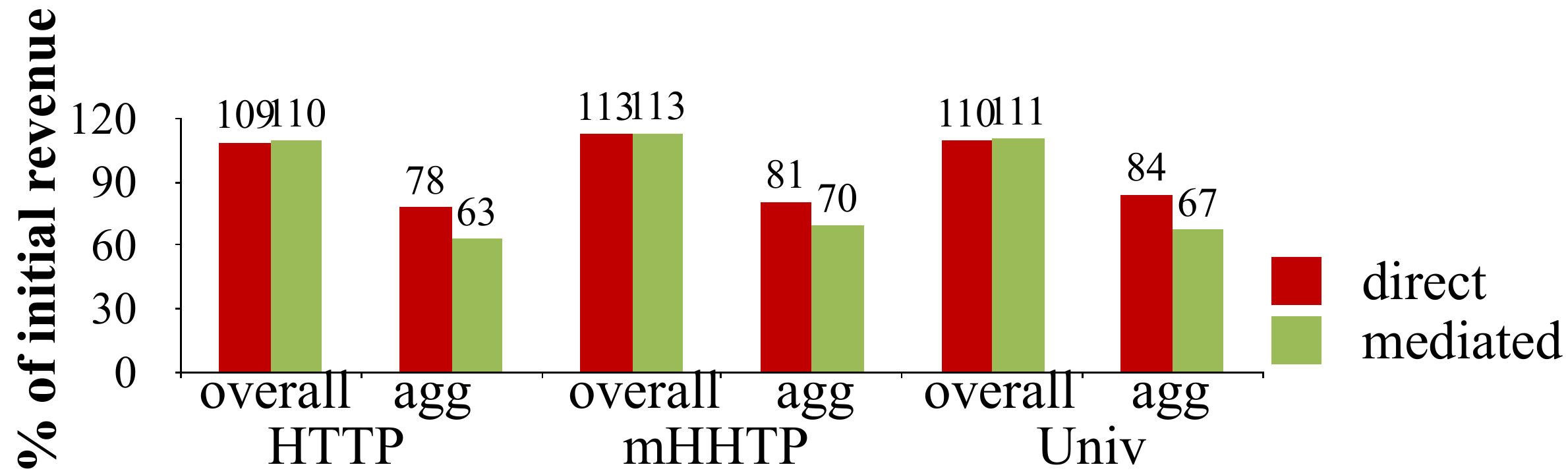}
\vspace{-0.1in}
\caption{Total revenue normalized to initial revenue overall and for aggregators.}
\vspace{-0.2in}
\label{fig:totalrev}
\end{figure}

\myparab{i) Markets increase overall revenue by 9-12\%}. Fig.~\ref{fig:totalrev} shows revenue  at the end of 
the simulation normalized to initial utility, both overall and for aggregators. There is no significant difference between 
the direct and mediated markets in terms of overall revenue increase, with both increasing overall revenue between 9-12\%. 



\begin{figure}[t]
\centering
\vspace{-1.1cm}
\includegraphics[width=0.4\textwidth ]{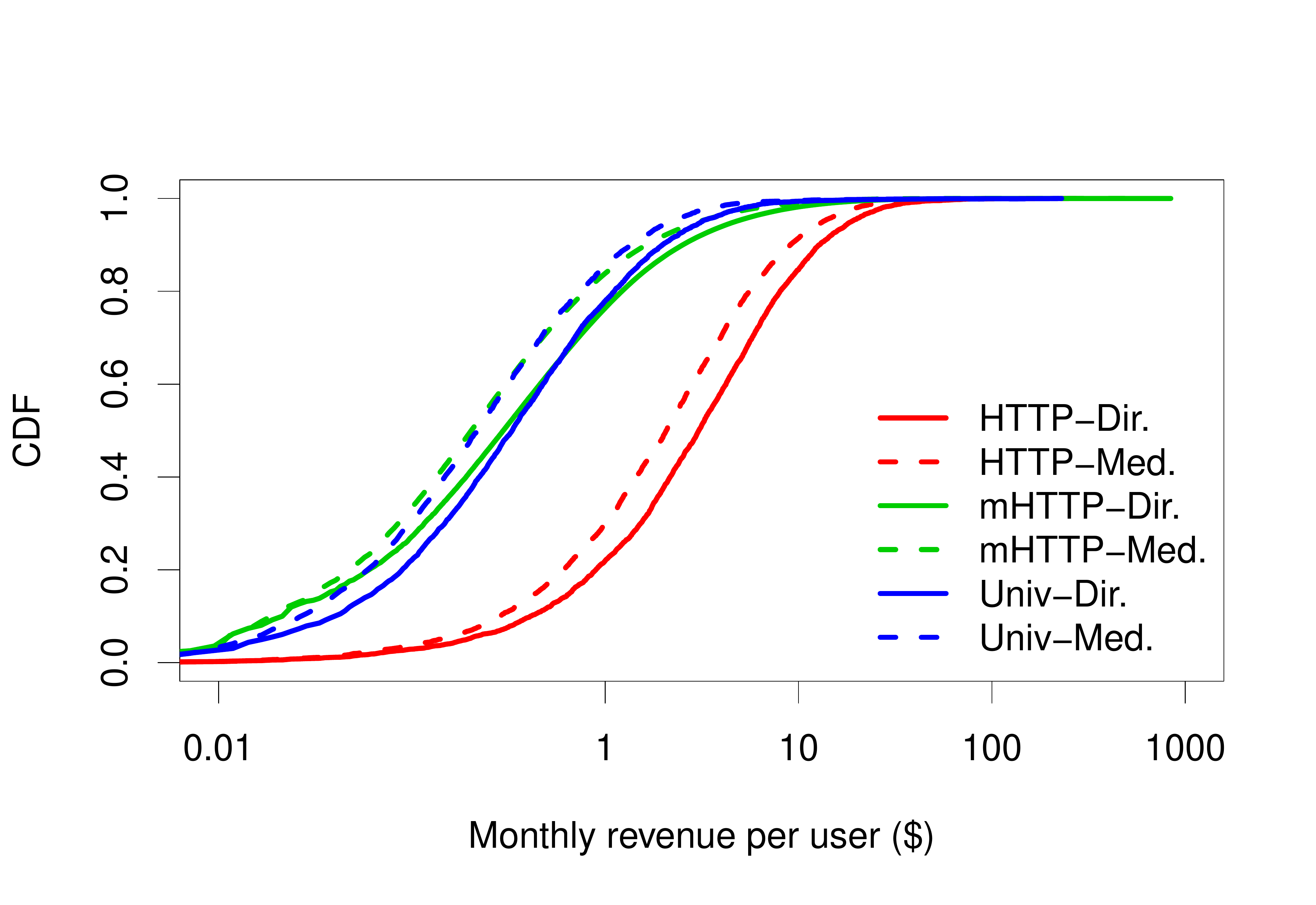}
\vspace{-0.2in}
\caption{Monthly revenue for users that participate in the direct and mediated market.}
\vspace{-0.1in}
\label{fig:userrev}
\end{figure}

\myparab{ii) Direct markets increase revenues for users.} Fig.~\ref{fig:userrev} shows projected monthly revenue for users in the 
direct and mediated markets. Users derive more revenue in 
the direct market since revenue is not shared with the trusted third party. The median monthly revenue in a direct market is 50\% higher than in a mediated one. 
\AMdel{We note here that these numbers should be viewed relatively and absolute values may vary in practice.}

\myparab{iii) User revenue is highly correlated with the number of aggregators.} Monthly revenue for users in the HTTP dataset is signficantly 
higher than in the university or mHTTP datasets. This stems from a high degree of correlation between aggregators the user comes in 
contact with and their revenue (correlation coefficients of 0.6-0.9). Indeed, the average number of 
aggregators per user is 43 in HTTP but only 5 in mHTTP and 9 in the university dataset.  The higher 
number of aggregators per user in HTTP can be due to multiple users sharing a connection (recall this is a residential broadband trace). 
\AMdel{We do not have enough additional information to investigate this.}
Next we consider how the market impacts revenue, overall and for aggregators.

\myparab{iv) Aggregator revenue decreases.} Revenue for aggregators decreases by 16-37\% as compared to today's status-quo after information market adoption. This result contrasts with the prediction seen in Sec.~\ref{sec:econmarket} that, for a \emph{single} transaction, aggregators always benefit from {\system}. This emerges when the system is analyzed at scale because a user interacts with multiple aggregators. If it joins an information market to obtain revenue from a transaction with an aggregator (typically, one that has high $r_{a}(u)$) part of the consequence is that tracking is blocked and this negatively affects another aggregator revenue (typically, one with low $r_{a}(u)$). However, if we compare these revenues to what would happen if users, publishers, or a regulation were to block tracking with DNT~\cite{dnt}, estimated in \cite{Gill:2013} to drop ad revenues by up to 75\%, deploying an information market allows aggregators to recover from this loss. It also addresses privacy concerns of the users, as information is obtained legally and transparently.

%% file: relwork.tex
\section{Related Work}
\label{sec:related}
The work presented in this paper lies at the intersection of online advertisements, privacy, and economics. 



\myparab{Privacy preserving advertisements and analytics.}

\hspace{-0.13in}There has been recent interest in solutions that protect users' privacy while still enabling aggregators to provide targeted advertisements, perform analytics
and personalized services \cite{Juels, Hub, Guha:2011kn, Toubiana:2010tm, Chen:2013, repriv, Riederer:2011}. 
Privad~\cite{Saikat:2011tt}, Adnostic~\cite{Toubiana:2010tm} and RePriv~\cite{repriv} are 
browser based systems that block access by 3rd parties to varying degrees and
enable targeted ads to be shown to the user, without leaking personal information. 
Information markets~\cite{Riederer:2011, datacoup, Laudon} also address the concerns of aggregators to collect information,
while allowing users to choose what information about them can be released for suitable
economic compensation. Our goal in this paper is to study the viability and effectiveness of such economically driven solutions
by designing, implementing and testing the solution in the wild with real users. 

\myparab{Privacy and economics.} Our work is closely tied to work which considers personal
information, through the lens of economics~\cite{Reznichenko:2011tx, evans, Ghosh:2011jy, Kleinberg:2001vz}.
Both Reznichenko \etal~\cite{Reznichenko:2011tx} and Ghosh \etal~\cite{Ghosh:2011jy} study and design auction mechanisms; the former deals with ad-space auctions and the latter deals with direct personal 
information (similar to~\cite{Riederer:2011}). Our work is different, in that we propose and implement a system that
combines privacy and performance, and study the \emph{implications}
of selling information (via a direct or mediated market) to various parties involved at a large scale using Web traces.



%% file: conclusions.tex
\section{Conclusions}
\label{sec:conc}
Can a system for online privacy be mutually beneficial to users and online advertisers? In this paper we take a first step towards answering this question by designing an information market focusing on 3rd-party tracking. We discussed the requirements for an information market, and designed and implemented our system to provide privacy, Web performance, and revenue to users. We evaluate feasibility of the market using a one-month field study with 63 users and observed that users actively engage with the system and are willing to sell parts of their browsing data, however these disclosures do not necessarily match up with their stated privacy concerns. Users also declared that using an information market affected their privacy attitudes.

We then proposed a model to capture the effects of information markets on advertising revenue and investigated their viability. Further, we considered the system at scale using traces containing billions of Web requests made by millions of users, to understand economic ramifications of an information market at scale and showed that it can be profitable to all parties. We found that advertising revenue increases by 9-12\% overall when all players, i.e., users and data aggregators, cooperate -- an observation that is at odds with most current beliefs about privacy preserving techniques. 

%% file: acknowledgment.tex
\section{Acknowledgment}

The authors would like to thank Dina Papagiannaki for advice and multiple suggestions while this experiment and analysis was conducted. This material is based upon work supported by the National Science Foundation under Grant No.1254035 and 1514437.

%% file: appendixData.tex
\section{Auction mechanism}
\label{app:auction}
\AMnote{We only mention this mechanism enforces truth telling, but do not explain why that's the case. We don't necessarily have to if this has been shown in the literature and it's a know fact. I just wasn't sure.}

Here we explain the exponential auction mechanism~\cite{exp-mechanism} that has the truth telling properties, and has been shown to be close to optimal in terms of revenue for the seller. Note that users are independent from each other and auctions are run per user.

We denote the user as $u$ and the set of aggregators by $\setA = \{a_0, a_1, ..., a_m\}$. The good being sold on the market is access to the user, \ie the ability to track the user on her whitelisted sites and serve her targeted ads for a given time interval. A user's browsing behavior can be sold to multiple aggregators with no marginal cost of reproduction, hence the market can be thought of as having an unlimited supply. Intuitively, aggregator $a$ should be willing to pay to access $u$ as long as the price to acquire her is smaller than the additional revenue, $v_{a}$, gained by having access to $u$. \AMdel{Extensions for an aggregator to buy exclusive access can be included although beyond the scope of this paper.} 




In the auction, we assume that each aggregator $a$ in $\setA$ who is interested in accessing $u$ bids a maximum price $p_{a}(u)$ that it is willing to pay for access to user $u$.
Assuming the final winning bid value is $p(u)$, every winning bidder, \ie bidders with bids higher than this value, pay $p(u)$, hence the total revenue of $u$ is given by:

\vspace{-0.2mm}
\[
 R\left[ (p_{a}(u)_{a\in\setA}, p(u) \right] = \sum_{a\in\setA} p(u) \times \ind{p(u)\leq p_{a}(u) }
 \;.
 \]
 \vspace{-0.1in}

When $p(u)>\max_{a\in\setA} p_{a}(u)$, the revenue will be zero, as the price exceeds what aggregators are willing to pay. We wish to choose $p(u)$ to maximize this sum; all bids higher than this value are considered winners
and hence are given access to the user. The winners pay $p(u)$. 

Following \cite{exp-mechanism} we first assign an initial value to $p(u)$ according to a measure $\nu(p(u))$ on $\Real$ and then re-weigh this measure to choose the actual price used. To re-weigh, we use an exponential function that puts more weight on high value of $R$, according to a parameter $\varepsilon>0$. 
PDF of the chosen price to track a given user is given by:
 \[
P(p(u)) = 
 \frac{\exp\left( \varepsilon R\left[ (p_{a}(u))_{a\in\setA}, p(u) \right] \right) \nu(p) }{ \int_{0}^{\infty}
 \exp\left(\varepsilon R\left[ (p_{a}(u))_{a\in\setA}, s \right]\right) \nu(s) ds }
 \]
A standard approach is then to choose the initial distribution of $p(u)$ according to the Lebesgue measure on $\Real$, such that $\nu(p(u))=1$.
 
By using $\varepsilon$, we have added noise around the value maximizing the revenue, given the set of bids. Although it seems counter-intuitive to use a suboptimal price, ~\cite{exp-mechanism} shows that this (1) prevents any bidder from winning more than a factor $\exp(\varepsilon)$ when cheating and 
(2) still reaches a revenue that is within a good bound of the optimal value if the number of aggregators is large. The expected revenue is at least $OPT - 3 \frac{\ln(e + OPT \epsilon^2 m ) }{ \epsilon}$, where $OPT$ denotes the optimal revenue and $m$ is the number of buyer aggregators in the optimal case.
\AMdel{Thus, although the randomization causes revenue from a given set of bids to be lower, truthful bidding means the set of bids will be higher, ending up with better revenue than if we allowed bidders to cheat.}

\section{Valuing user data}
\label{apx:eval}
The goal is to estimate, for a given period [0, T], the advertisement revenue that can be generated from a user based on her whitelist and impressions that she generates (frequency of visits). We want the model to be simple, and intuitively monotone in the number of impressions and size of whitelists. We rely on keywords associated with websites to estimate the ad-placing strategy maximizing the total clicks generated. We assume: 1) the period is short enough so that any advertiser would like to receive a single click from the same user. Equivalently, the user would never click twice on an ad by the same advertiser in the given period. In both cases, playing an ad after is was clicked creates no revenue. 2) the period is short enough that a visit to a website at any time denotes a topic of relevance to that user for the entire period and 3) the period is long enough (or the system loses memory sufficiently quickly) so that any action (ads shown, clicks, revenue generated) from the past period is irrelevant.

\newcommand{\pclic}{\pi^{\textrm{click}}}
\newcommand{\cpc}{\textrm{\texttt{CPC}}}

We denote users by $u$, publisher or website by $j$, and advertisers by $a$. Each website is associated with one or several keywords that we denote by $\kappa(j)$. Similarly, each advertiser has a set of relevant keywords denoted by $\kappa(a)$. Whenever a user visits a website $j$ and reveals that information, she effectively discloses the associated keywords. We denote by $\kappa(u) = \cup_{j \textrm{visited by} u} \kappa(j)$ the set of all these keywords and are all relevant to target ads to $u$.
We assume that when $u$ visits $j$ and sees an ad from advertiser $a$, she decides independently with probability $\pclic (u, j, a)$ to click on the ad, unless it has clicked on it already during that period. For simplicity we start with a constant probability, but in general this depends on the website, the advertiser (typically through the associated keyword), and the user. For instance, one may imagine that a user is more likely to click for an advertiser associated with a keyword that is associated with many sites she visits.
Advertiser $a$ can serve an ad to a user $u$ if and only if they share one relevant keywords $\kappa(u)\cap\kappa(a)\neq\emptyset$. It typically obtains this slot through a bidding process. We will assume for simplicity that all advertisers' bids are equal to the average cost per click (CPC) for a keyword that they have in common, and we denote this value by $\cpc(k)$. If this intersection has multiple keywords, then the advertiser will typically bid with the highest keywords as it denotes higher interest in the user. These values are estimated using Google AdWords keyword planner tool.

During the period, a user generates $N(u,j)$ impressions on a website $j$. Out of the total number of impressions created, the system decides to display a number of ads associated with $a$, $n_a$. The probability that all of these fail to generate a click is $(1-\pclic(u, j, a))^{n_a}$. Hence the optimal expected value generated by clicks, depending on which ads are played, can be found as the solution of the following optimization problem, which can be solved by a simple greedy algorithm:
\[
\begin{disarray}[c]{c}
\max \sum_{k\in\kappa(u)} \cpc(k) \sum_{a \text{ such that } k(a)=k} \left(1 - \pclic(u, j, a) \right)^{n_a} \\
\midwor{such that}
\sum_a n_a = \sum_j N(u,j) = N(u)
\end{disarray}
\]

%% file: appendixShapley.tex
\section{Shapley value overview}
\label{apx:shapley}
In a cooperative game, the set of players is denoted as $\setN$. We call any subset $\setS \subseteq N$ a \emph{coalition} of players. 
For each coalition $\setS$, we denote by $V(\setS)$ the \emph{worth function}, which measures the total revenue produced 
as a result of the coalition $\setS$. 



We define the \emph{marginal contribution} of player $i$ to a coalition ${\setS} \subseteq {\setN}\backslash \{i\}$ as
$\Delta_i({\setS},V) = V({\setS}\cup \{i\})-V({\setS})$. 
Note that the contribution of a player only depends on the worth function $V({\setS})$.

Shapley value determines 
how the total worth of the coalition, captured by $V(\setS)$, should be shared among the players in $\setS$. More specifically, the Shapley value of player $i$ is denoted by
 $\varphi_i\left( \setS,V \right)$
and is uniquely defined by the following three axioms:

{\bf  Axiom 1:} (Efficiency) $\sum_{i\in {\setS}}\varphi_{i}({\setS},V)=V({\setS})$. This ensures that revenue assigned to the players is the total revenue created by the coalition.

{\bf  Axiom 2:} (Symmetry) 
If for all ${\setS' \subseteq {\setS}} \backslash \{i,j\}$, $\ds V({\setS'} \cup \{i\})=V({\setS'} \cup \{j\})$, then $\varphi_{i}({\setS},V)=\varphi_{j}({\setS},V)$. This means that two players who contribute the same amount to revenue receive an equal share of the revenue created.

{\bf  Axiom 3:} (Fairness/Balanced Contribution) 
For any $i, j\in \setS$, $j$'s contribution to $i$ equals $i$'s contribution to $j$, or, in other words 
$\varphi_{i}({\setS},V)-\varphi_{i}({\setS} \backslash \{j\},V)=\varphi_{j}({\setS},V)-\varphi_{j}({\setS} \backslash \{i\},V)$. This addresses fairness between any pair of players.

Based on the axioms above, one can show that the {\it Shapley value}
$\varphi$ can be computed as follows~\cite{shapley-orig}:
\begin{equation}
\forall \ i\in \setS \vf
\varphi_i({\setS}, V) = \frac{1}{|\setS|!} \sum_{\pi \in \Pi}
\Delta_i(S(\pi,i), V) \hspace{5mm} 
\label{equation:Shapley}
\end{equation} where $\Pi$ is the set of all $|\setS|!$ orderings of $\setS$ 
and $S(\pi,i)$ is the set of players preceding $i$ in the ordering $\pi$.

\noindent The Shapley value of a player $i$ can thus be
interpreted as the \emph{expected} marginal contribution 
$\Delta_i(\setS',V)$ where $\setS'$ is the set of players in $\setS$ preceding $i$ in a uniformly 
distributed random ordering of $\setS$.